\documentclass[aps,prl,twocolumn,showpacs]{revtex4}

\usepackage{amsfonts}
\usepackage{amssymb}
\usepackage{amsmath}
\usepackage[dvips]{graphicx}
\usepackage[dvips]{psfrag}

\newcommand{\bra}{\langle}
\newcommand{\ket}{\rangle}

\newcommand{\bv}[1]{{\boldsymbol #1}}
\newcommand{\e}{{\rm e}}

\newcommand{\Js}{J_{\rm s}}
\newcommand{\js}{j_{\rm s}}
\newcommand{\Is}{I_{\rm s}}

\begin{document}

\title{Variational formula for experimental determination 
of high-order correlations of current 
fluctuations in driven systems}

\author{Takahiro Nemoto and Shin-ichi Sasa}
\affiliation{Department of Basic Science,
The University of Tokyo, Tokyo, 153-8902, Japan}
\date{\today}

\begin{abstract}
For Brownian motion of a single particle subject to 
a tilted periodic potential on a ring, we propose a 
formula for experimentally determining the cumulant 
generating function of time-averaged current without 
measurements of current fluctuations. 
We first derive this formula phenomenologically on 
the basis of two key relations: a fluctuation relation
associated with Onsager's principle of the least energy 
dissipation in a sufficiently local region and  an 
additivity relation by which spatially inhomogeneous 
fluctuations can be properly considered. We then
derive the formula without any phenomenological
assumptions. We also demonstrate its practical 
advantage by numerical experiments.
\end{abstract}

\pacs{05.40.-a, 05.70.Ln, 02.50.Ey}

\maketitle


\paragraph{Introduction:}


Transportation properties in  linear response regime
out of equilibrium are determined by the second-order
correlation of time-averaged current fluctuation in 
equilibrium \cite{Onsager}. As an extension of this well-established 
formula, a nonlinear response formula was derived \cite{KG},
where stronger non-linearity arises when the third and higher 
order correlations are more relevant. 
Thus, toward a systematic understanding of strongly 
nonlinear transportation,  a 
universal law concerning high-order correlations
must be considered. 


In general, let $\hat I$ be a time-averaged current during
a time interval $\tau$. Its statistical
property is characterized by a cumulant generating
function $G(h)$, which is  defined by
\begin{equation}
e^{\tau G(h)}=\bra \e^{\tau h  \hat I} \ket
\label{def:lambda}
\end{equation}
for sufficiently large $\tau$. In recent years, 
$G(h)$ has been studied theoretically for several 
models \cite{BD,BDGJL,
phase-transition1,phase-transition2,phase-transition4,
Cedric,Thomas,Cristian}, partly motivated by its
symmetry property called fluctuation theorem \cite{FT}.
However, it is difficult to experimentally measure 
$G(h)$ when the distribution 
of $\hat I$ deviates far from the Gaussian. Since physically important 
quantities should be obtained experimentally, there should 
be a formula for making $G(h)$ measurable. 


Here, let us recall Einstein's  theory for macroscopic 
fluctuation of thermodynamic variables in equilibrium.
As an example, the cumulant
generating function of space-averaged magnetization $\hat m$
in  magnetic materials without a magnetic field is equivalent 
to the Gibbs free energy density as a function of the magnetic 
field, and the thermodynamic functions can be 
determined by measurements of heat capacity and 
susceptibility for any magnetic field and temperature.  
This implies that 
the cumulant generating function of $\hat m$ 
is obtained 
without measurements of fluctuations.
A universal idea behind Einstein's theory is that
when a state is determined by an extreme condition 
of a function, 
fluctuation properties at  more microscopic scales
are described by the variational function. 
The relation between the least action principle 
in classical mechanics and the path-integral 
formulation of quantum mechanics follows the
same idea. We thus conjecture that  the cumulant
generating function $G(h)$ of $\hat I$
might be identified with a variational function associated
with a  variational principle that determines a steady state.


In this Letter, 
we study Brownian motion of a single particle 
subject to a tilted periodic potential on a ring.
Starting from the analysis of a  simple example 
based on the least energy dissipation principle \cite{Onsager}, 
we  propose a phenomenological formula for 
experimentally determining the cumulant generating 
function of time-averaged current.
We then derive the formula theoretically and 
measure $G(h)$ in  numerical experiments.

\paragraph{Basic idea:}


We investigate an  electric circuit where 
three resistances, $R_0$, $R_1$ and $R_2$, 
are connected in series under voltage $V_0$.
In particular, in the limiting case that
$V_0 \to \infty $ and $R_0 \to \infty$ with
$I=V_0/R_0$ fixed, the system is assumed
to be in the constant-current environment.
Furthermore, we can impose the constraint 
that the voltages over the 
resistance $R_1$  and $R_2$ are fixed 
as $V_1$ and $V_2$, respectively. The total 
energy dissipation ratio in this system is 
given by 
\begin{equation}
\Phi(V_1,V_2|I)=\frac{V_1^2}{R_1}+\frac{V_2^2}{R_2}-2(V_1+V_2)I,
\end{equation}
where we ignore a contribution independent of $V_1$ 
and $V_2$.  When we remove the voltage constraint,
the values of $V_1$ and $V_2$ are determined by 
the condition that the current is equal to $I$.
Here, these values minimize $\Phi(V_1,V_2|I)$. 
This is an example of the least energy 
dissipation principle.


We next investigate another electric circuit, 
where two resistances, $R_1$ and $R_2$, are connected 
in series and the resistance $R_0$ is  connected in
a parallel way under the total current $I_0$.
In the limiting case that $I_0 \to \infty $ 
and $R_0 \to 0$ with $V=R_0 I_0$ fixed, the 
system is assumed to be in the constant-voltage 
environment. We can impose the constraint that 
the current passing the resistances $R_1$  and $R_2$ 
is fixed as $I$. 
The total energy dissipation ratio in this system 
is given by 
\begin{equation}
\Psi(I|V)=(R_1+R_2)I^2-2VI,
\end{equation}
where we ignore a contribution independent of $I$.
Without the current constraint, the value of $I$
is determined by the condition that the voltage 
is equal to $V$. Here, this value minimizes 
$\Psi(I|V)$. This is another example of the 
least energy dissipation principle.


Now, a variational principle is expected to be 
related to the description of fluctuation 
properties  as discussed in {\it Introduction}.
For example, the probability of time-averaged
current fluctuations in the constant-voltage 
environment is given as 
\begin{equation}
{\rm Prob}(I|V)\simeq e^{-\tau \frac{1}{4T}[\Psi(I|V)+c_0]},
\label{onsager}
\end{equation}
where $c_0$ is the normalization constant 
and the coefficient $4T$ 
respects the fluctuation-dissipation relation and 
the Boltzmann constant is set to unity. 
Then, the cumulant generating function $G(h|V) $ 
is determined by
\begin{equation}
G(h|V)=-\frac{1}{4T}\min_{I}[\Psi(I|V) -4ThI+c_0].
\label{GhV}
\end{equation}
The condition $G(h=0|V)=0$
gives $c_0=-\Psi(\Is|V)$, where $\Is$ 
represents the steady current.
Although (\ref{GhV}) is a simple mathematical expression,
we here attempt to rewrite it so as to be measured
experimentally. The idea is to consider a modified system
in which an extra voltage $v$ in addition to $V$ is applied.
We define  $v_{i*}\equiv vR_i/(R_1+R_2)$ which represents
the allocation of the extra voltage $v$ in the circuit.
We also denote by $I_{\rm s}^{v}$ 
the steady current for the modified system.
By using these quantities, we can express 
$\Psi(I|V)- 4ThI+c_0$ as 
\begin{equation}
 \Phi (v_{1*}, v_{2*}|I_{\rm s}^{v})
+\sum_{i=1}^{2}R_i(I-I_{\rm s}^{v})^2 - 2I(2Th - v) .
\label{pp}
\end{equation}
By choosing  $v = 2Th$ and minimizing 
(\ref{pp}) in terms of $I$, 
we obtain $G(h|V)=- \Phi (v_{1*}, v_{2*}|I_{\rm s}^{v})/(4T)$.
The variational principle in the first example leads us to
rewrite it as an additive form
\begin{equation}
G(h|V)=
-\frac{1}{4T}\min_{v_1+v_2=2Th}\sum_{i=1}^2
\left[ \frac{v_i^2}{R_i}-2I_{\rm s}^{v}v_i
\right].
\label{form}
\end{equation}
This additivity relation is the basic formula which we extend so as to apply
more general cases. Since the voltages can be controlled
experimentally, the right-hand side of (\ref{form}) is
obtained experimentally. Although the formula is derived for
the simple system, the result is quite suggestive. 
The assumption
for the formula is that the least energy dissipation principle
and the fluctuation-dissipation relation hold locally in space.
Although this assumption is not always valid, many non-equilibrium 
systems are included in this class. 
Below we apply the relation (\ref{form})
to the Brownian motion out of equilibrium.

\paragraph{Main Result:}


The system we study consists of a single Brownian particle 
on a ring with a size of $L$. Its motion is described by a Langevin 
equation 
\begin{equation}
\gamma \dot{x}=f-\frac{\partial U}{\partial x}+\xi,
\label{Lan}
\end{equation}
where $\dot x\equiv dx/dt$ and $\xi$ is the noise 
satisfying $\bra \xi(t)\xi(t')\ket=2T\gamma \delta(t-t')$.
$\gamma$ represents a friction constant, $f$ a uniform 
driving force, and $U(x)$ a periodic potential. 
The external force acting on the particle is written as
$F(x)=f-\partial U(x)/\partial x$. 
It should be noted that 
the Langevin equation (\ref{Lan}) corresponds to 
experimental systems that have been studied with
the aim of testing new ideas in 
non-equilibrium statistical mechanics \cite{Bechinger,Ciliberto,Toyabe}.
We are particularly concerned with 
the cumulant generating function  $G(h)$ 
defined in (\ref{def:lambda}) with $\hat I$ replaced by 
\begin{equation}
\hat v=\frac{1}{\tau } \int_0^\tau dt \frac{dx}{dt}.
\label{hatv}
\end{equation}
Throughout this Letter, the boldface font is used 
to indicate a function of $x$. 


Here, in order to find a formula similar to (\ref{form}), 
we apply an extra force $u(x)$ in addition to $F(x)$
with an interpretation that $F(x) \Delta x$ and $u(x) \Delta x$ 
correspond to the voltage and the excess voltage over the 
interval $[x,x+\Delta x]$, respectively.
We consider an empirical probability
density $\rho(x)$ and its current $j(x)$ obtained by 
measurement during a finite time interval $\tau$. 
That is, $\rho(x) \equiv \int_0^\tau dt \delta(x-x(t))/\tau$
and $j(x) \equiv \int_0^\tau dt \delta(x-x(t)) \circ\dot x/\tau$,
where $\circ $ is the Stratonovich rule of the multiplication.
In the limit $\tau \to \infty$,
$j(x) \to J_{\rm s}^{\bv{u}}$ and $\rho(x) \to \rho_{\rm s}^{\bv{u}}(x)$,
where the steady probability density $\rho_{\rm s}^{\bv{u}}(x)$
and the spatially homogeneous 
current $J_{\rm s}^{\bv{u}}$ for this modified
system with the extra force $\bv{u}$ are determined by
\begin{eqnarray}
J_{\rm s}^{\bv{u}}=\frac{1}{\gamma}\rho_{\rm s}^{\bv{u}} (F+u)
-\frac{T}{\gamma} \partial_x \rho_{\rm s}^{\bv{u}}.
\label{Js-def}
\end{eqnarray}
This shows that the resistance in the interval $[x,x+\Delta x]$ 
corresponds to $\gamma (\Delta x)/\rho_{\rm s}^{\bv{u}}(x)$. 
Finally, since the current $\hat I$ corresponds to 
$\int_0^L dx j(x)/L=\hat v/L$, $h$ in the formula for 
the time averaged  current is replaced by $hL$ 
for the case of cumulant generating  function $G(h)$ 
for $\hat v$. That is, the constraint condition
$v_1+v_2=2Th$ becomes
\begin{equation}
\int_0^L dx u(x)=2ThL.
\label{constraint}
\end{equation}
From these correspondences, we heuristically derive
\begin{eqnarray}
G(h)
=-\frac{1}{4T} 
\min_{\bv{u};\int_0^L dx u(x)=2ThL} {\cal G}(\bv{u}),
\label{final-1}
\end{eqnarray}
with 
\begin{eqnarray}
{\cal G}(\bv{u})=
\int_0^L dx 
\left[
\frac{\rho_{\rm s}^{\bv{u}}(x) }{\gamma}
u(x)^2-2 J_{\rm s}^{\bv{u}}u(x)
\right].
\label{final-2}
\end{eqnarray} 
That is, $G(h)$  is proportional to the  minimum energy dissipation
of the extra force $u(x)$ when the system is assumed to be
in constant-current environment of the current $J_{\rm s}^{\bv{u}}$
with $\int_0^L dx u(x)=2T hL$. The formula (\ref{final-1}) with
(\ref{final-2}) is  the main claim of this Letter. 

\begin{figure}[tbh]
\includegraphics[width=4cm]{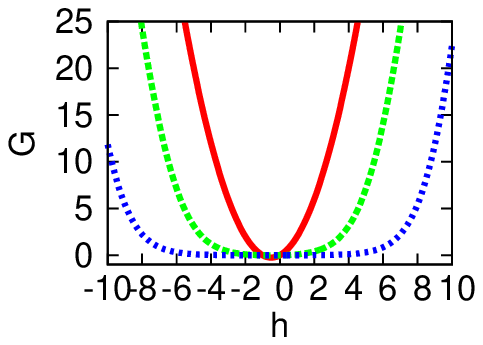}
\includegraphics[width=4cm]{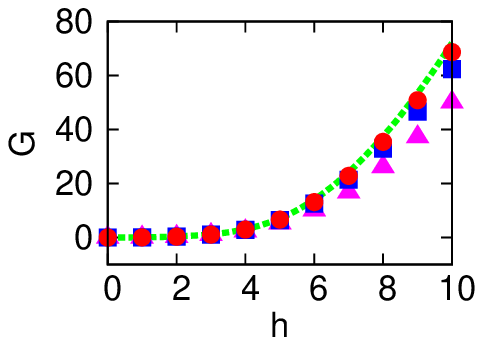}
\caption{(color online). $G(h)$ for 
the Langevin equation (\ref{Lan}) with 
$U(x)=U_0\cos (2 \pi x/L)$. Quantities are
converted to dimensionless forms by setting 
$\gamma=T=L=1$. We fix $f=1$. 
(Left:)  Numerical calculation of $K=2TG(h)$ 
in (\ref{K-def}) for $U_0=0$ (red solid line), 
$3$ (green dashed line), and 
$5$ (blue dotted line).
(Right:) Experimental measurement for the system with
$U_0=3$. We assume to know $T=\gamma=1$ and determined
$G(h)$ experimentally from trajectories $(x(t))_{t=0}^\tau$
according to the method described
in the text. $N=5$ and $\tau=250$ (triangle symbols), $1000$ (square
symbols), and $4000$ (circle symbols). The error-bars are within the
symbols. The dashed line is the same as that in the left-side.}
\label{fig1}
\end{figure}

We determine the optimal force $u_{\rm opt}(x)$ 
that satisfies 
${\cal G}(\bv{u}_{\rm opt})
=\inf_{\bv{u}}{\cal G}(\bv{u})$
with the condition  (\ref{constraint}).
By calculating  $\delta {\cal G} \equiv 
{\cal G}(\bv{u}+\delta \bv{u})-{\cal G}(\bv{u})$ 
explicitly, we find that $\delta {\cal G}=0$
leads to 
\begin{eqnarray}
K = \frac{1}{\gamma}
\left( F+ \frac{u_{\rm opt}}{2} \right)u_{\rm opt}
+\frac{T}{\gamma}\partial_x u_{\rm opt}, 
\label{K-def}
\end{eqnarray}
where $K$ is a constant. Differentiating the both-sides 
with respect to $x$, we obtain $u_{\rm opt}(x)$ under
the periodic boundary condition. The substitution of 
the obtained result $u_{\rm opt}$ into (\ref{K-def})
leads to $K$. This constant $K$ is significant.
Indeed, by multiplying $\rho_{\rm s}^{\bv{u}_{\rm opt}}$
to (\ref{K-def}) and $u_{\rm opt}$ to (\ref{Js-def}) and 
by considering the difference between them, 
we obtain $K=2T G(h)$. 
Furthermore, for the non-linear 
differential equation (\ref{K-def}) with $K=2TG(h)$,
we perform a Cole-Hopf transformation defined by
$u_{\rm opt}(x)=2T(h+\partial  \log \psi(x) / \partial x )$,
where $\psi (x) >0$. We then derive a linear eigenvalue
equation ${\cal L}_h \psi= G(h)\psi$, where the operator
${\cal L}_h$ is defined by
\begin{equation}
{\cal L}_h \cdot 
\equiv \frac{F}{\gamma}(\partial_x+h) \cdot
+\frac{T}{\gamma}(\partial_x +h)^2\cdot \ .
\label{calL}
\end{equation}
From the positivity of  $\psi$, $G(h)$ turns out to be 
equal to the maximum eigenvalue $\lambda_{\rm max}$
of the operator ${\cal L}_h$ and $u_{\rm opt}$ is related to
the corresponding eigenfunction. 
Examples of 
$G(h)$ thus determined uniquely are displayed in the left-side
of Fig. \ref{fig1}.


Now, we prove (\ref{final-1}) 
without employing any phenomenological assumptions.
For a stochastic variable $\hat y(t) \equiv t \hat v(t)
=\int_0^t ds dx(s)/ds$, we denote by $P(y,t)$ the 
probability density that $\hat y$ takes the value 
$y$ at time $t$. We then 
have $\exp(\tau G(h))=\int dy P(y,\tau)\exp(hy)$.
Here, the function  $Q(y,t)\equiv P(y,t)\exp(hy)$
obeys $\partial_t Q={\cal M}_h Q$, where
\begin{equation}
{\cal M}_h \cdot 
=
-\frac{1}{\gamma}(\partial_y-h)(F(y) \cdot \ )
+\frac{T}{\gamma}(\partial_y -h)^2\cdot. 
\label{calM}
\end{equation} 
We denote the maximum eigenvalue of the operator ${\cal M}_h$ by
$\mu_{\rm max}$. Since $Q(y,t) \simeq e^{\mu_{\rm max} t}$
for sufficiently large $t$, $G(h)$ is equal to $\mu_{\rm max}$.
(See a related result in Ref. \cite{Thomas}.)
By comparing (\ref{calL}) and (\ref{calM}),
we find  ${\cal L}_h^\dagger={\cal M}_h$. Thus,
$\lambda_{\rm max}=\mu_{\rm max}$. Since 
the eigenvalue equation ${\cal L}_h \psi= G(h)\psi$
is equivalent to the variational form  (\ref{final-1}) 
with (\ref{final-2}), we end the proof.
This method of derivation of the formula
can also be applied to Langevin 
equations in any dimensions. See Ref. \cite{Nemoto-Sasa}
for details. 

The formula (\ref{final-1})  is closely 
related to the so-called additivity principle
\cite{BD,BDGJL, phase-transition1,phase-transition2,
phase-transition4}. We mention  the relation explicitly. 
Let us consider the probability density of 
$\hat J=\hat v/L$,
which  is expected to possess a large deviation property
\begin{equation}
{\rm Prob}(\hat J=J) \simeq \e^{-\tau \Gamma(J)}.
\end{equation}
Through mapping from (\ref{Lan}) to fluctuating hydrodynamics 
\cite{Dean}, we can apply the additivity principle to the 
continuum description of the fluctuating density field. 
The result is 
\begin{equation}
\Gamma(J)=
\min_{\bv{\rho}}
\int_0^L dx \frac{(J-\js^{0}(x;\bv{\rho}))^2}
{4T \rho(x)/\gamma},
\label{add}
\end{equation}
where 
$\js^{\bv{u}}(x;\bv{\rho})
\equiv\rho (F(x)+u(x))/\gamma-T \partial_x \rho/\gamma$
for any $u(x)$. 
Since $G(h)$ is connected to $\Gamma(J)$ as
$G(h)=\max_{J} [ hLJ - \Gamma(J)]$,
we obtain
\begin{equation}
G(h)=-\frac{1}{4T}\min_{J, \bv{\rho}}
\int_0^L dx \phi(x;J,\bv{\rho}),
\label{addgdef}
\end{equation}
where
$$
\phi(x)=\frac{\rho (x)}{\gamma} u(x)^2
-2 u(x) \js^{\bv{u}}(x;\bv{\rho})
+\frac{\gamma}{\rho(x)}[J-\js^{\bv{u}}(x;\bv{\rho})]^2,
$$
under the constraint condition (\ref{constraint}).
We make a variable transformation from $\rho$ to $u$ by 
$\partial_x \js^{\bv{u}}(x;\bv{\rho})=0$.
Explicitly, for a given $u$, we can determine $\Js^{\bv{u}}$ 
and $\rho_{\rm s}^{\bv{u}}(x)$ uniquely
as $\js^{\bv{u}}(x;\bv{\rho}_{\rm s}^{\bv{u}})
=\Js^{\bv{u}}$.
The minimization with respect to $J$ in (\ref{addgdef})
is achieved by $J=\Js^{\bv{u}}$,
and $\int_0^L dx \phi(x;J,\bv{\rho}_{\rm s}^{\bv{u}})$ 
becomes ${\cal G}(\bv{u})$. Thus, 
(\ref{addgdef}) leads to 
the formula (\ref{final-1}).

\paragraph*{Experiment:}


We consider the experimental measurement of $G(h)$.
In the standard method, $G(h)$ may be determined
from the measurements of cumulants. However,
high-order cumulants are  quite  
difficult  to be measured experimentally. 
Here, we claim that our formula can provide $G(h)$ 
experimentally if we can control the external 
force and know the value of the temperature 
$T$ and the friction constant $\gamma$. The method
is as follows.

We measure  a trajectory $(x(t))_{0 \le t \le \tau}$
for the system with an extra force $w(x)$.
From this data, we define 
\begin{equation}
\tilde {\cal G}_\tau(\bv{w})\equiv 
\frac{1}{\tau}\int_0^\tau dt
\left[
\frac{w(x(t))^2}{\gamma}-2 \dot x(t) \circ w (x(t))
\right].
\end{equation}
Note that $\tilde {\cal G}_\tau(\bv{w}) \to {\cal G}(\bv{w})$
in the limit $\tau \to \infty$. Thus, if we know the 
optimal force $u_{\rm opt}$, the approximate value
of $G(h)$ is obtained from $\tilde {\cal G}_\tau(\bv{w})$
by setting ${w}=u_{\rm opt}$. 
Here, we have an identity
\begin{equation}
{\cal G}(\bv{u}_{\rm opt})=
-4Th \Js^{\bv{w}}L
-\int_0^L dx \frac{\rho_{\rm s}^{\bv{w}}}{\gamma}
u_{\rm opt}(u_{\rm opt}-2w)
\label{ident}
\end{equation}
for any $\bv{w}$, which can be derived from the
calculation of ${\cal G}(\bv{u}_{\rm opt})-{\cal G}(\bv{w})$ 
with the aid of (\ref{Js-def}) and (\ref{K-def}) 
\cite{Nemoto-Sasa}.
We express $u_{\rm opt}(x)=2Th+\sum_{n=-N}^N 
a_n \phi_n(x)$, where $\phi_n(x)=\cos(2\pi n x/L)$ for $n \ge 0$,
$\phi_n(x)=\sin(2\pi n x/L)$ for $n < 0$, $a_0=0$, and 
$N$ is the truncation number of the approximation of the 
force $u_{\rm opt}(x)$. Since (\ref{ident}) holds
for any $\bv{w}$, we determine $(a_n)_{n=-N}^N$ by
preparing $2N+1$ forces $w^{(i)}(x)$, $-N \le i \le N$.
Concretely, by using the trajectory $x^{(i)}(t)$ for each 
$w^{(i)}(x)=5 T\phi_i(x)/L$, the right-hand side of 
(\ref{ident}) is approximated as
\begin{equation}
\tilde H_\tau(\bv{w}^{(i)})
\equiv \frac{1}{\tau}\int_0^\tau dt {\cal H}^{(i)}(t),
\end{equation}
where
\begin{eqnarray}
{\cal H}^{(i)}(t)
\equiv &- 4Th \dot x^{(i)}(t)
+2\frac{1}{\gamma} \sum_{n}\phi_n(x^{(i)}(t))w^{(i)}(x^{(i)}(t))a_n
\nonumber  
\\
& 
-\frac{1}{\gamma}\sum_{n,m}\phi_n(x^{(i)}(t))\phi_m(x^{(i)}(t)) 
a_n a_m.
\end{eqnarray}
Since $\tilde H_\infty(\bv{w}^{(i)})$ is independent of $\bv{w}^{(i)}$,
we consider the  following equation for $(a_n)_{n=-N}^N$:
\begin{equation}
\tilde H_\tau(\bv{w}^{(i)})=\tilde H_\tau(\bv{w}^{(i+1)}),
\label{eqforopt}
\end{equation}
where $-N\le i < N$. By solving (\ref{eqforopt}),
we obtain an approximation of $u_{\rm opt}(x)$. 
The approximation becomes more accurate for larger $N$ 
and larger $\tau$. In the right-side of Fig. \ref{fig1}, 
we display an example 
of the measurement of $G(h)$ in numerical experiments. 

\paragraph*{Concluding remarks:}


In this Letter, we have proposed the formula (\ref{final-1})
for a driven Brownian motion described by (\ref{Lan}).
We have demonstrated that $G(h)$ can be determined experimentally
from $T$, $\gamma$, and trajectories of the modified system
with an external force. Note that the formula  
yields an expression of the $k$-th order cumulant $C_k = \tau^{k-1}\bra 
\hat v^k \ket_{\rm c} $ \cite{Nemoto-Sasa}, 
where  $C_1 $ and $C_2 /2$ are equal to the average velocity and 
diffusion constant of the Brownian particle, respectively. 
The result reproduces the known expression for 
the diffusion constant \cite{Reimann} in the simplest manner.
Furthermore, (\ref{final-1}) 
can be extended to more general bulk-driven  
systems when we assume fluctuating hydrodynamics. 
We express $G(h)$ by a path-integral expression, 
and  if it is determined by the contribution of a saddle
configuration, the formula (\ref{final-1}) 
is valid, as for the case of (\ref{Lan}). 
However, there are many systems for which 
the condition is not satisfied,
as investigated in Refs. \cite{BDGJL,phase-transition1,
phase-transition2,phase-transition4}. (See also 
sections 3 and 4 in \cite{PTPS} for directly related 
discussions on the problem in this Letter.)
Future work will concern the determination of
the range of the application of the formula. 
We believe that our theory for the derivation 
provides a sound approach in studying this 
problem. We also wish 
to present  non-trivial predictions by employing 
the formulas for interacting particles systems. 
We hope that our formula will be studied 
theoretically, numerically, and experimentally, so that 
the novel nature of non-equilibrium fluctuation can be
uncovered. 


The authors thank F. van Wijland for
his fruitful comments on the earlier
version of this Letter. They also thank
T. Bodineau and H. Tasaki for discussions
for the theoretical derivation.
This work was supported 
by grants from the Ministry of Education, 
Culture, Sports, Science, and Technology 
of Japan, Nos. 21015005 and 22340109. 


\end{document}